\documentclass{article}

\usepackage{amsmath,amssymb,amsfonts}
\usepackage{algorithmic}
\usepackage{graphicx}
\usepackage{textcomp}
\usepackage{enumitem}
\usepackage{url}
\usepackage{xcolor}

\usepackage[ruled,vlined,linesnumbered]{algorithm2e}

\SetCommentSty{mycommfont}

\def \FF {{\mathbb F}}
\DeclareMathOperator*{\argmax}{arg\,max}
\DeclareMathOperator*{\argmin}{arg\,min}

\usepackage{authblk}
\title{Side-Channel Analysis of OpenVINO-based Neural Network Models}
\author[1]{Dirmanto Jap}
\author[2]{Jakub Breier}
\author[3]{Zdenko Lehock\'{y}}
\author[1]{Shivam Bhasin}
\author[3]{Xiaolu Hou}
\affil[1]{National Integrated Centre for Evaluation (NiCE) \\Nanyang Technological University, Singapore. \{djap,sbhasin\}@ntu.edu.sg}
\affil[2]{TTControl GmbH, Vienna, Austria. jbreier@jbreier.com}
\affil[3]{Slovak University of Technology, Bratislava, Slovakia. xlehocky@stuba.sk, houxiaolu.email@gmail.com}
\date{}                     
\begin{document}

\maketitle

\begin{abstract}
Embedded devices with neural network accelerators offer great versatility for their users, reducing the need to use cloud-based services.
At the same time, they introduce new security challenges in the area of hardware attacks, the most prominent being side-channel analysis (SCA).
It was shown that SCA can recover model parameters with a high accuracy, posing a threat to entities that wish to keep their models confidential.

In this paper, we explore the susceptibility of quantized models implemented in OpenVINO, an embedded framework for deploying neural networks on embedded and Edge devices.
We show that it is possible to recover model parameters with high precision, allowing the recovered model to perform very close to the original one.
Our experiments on GoogleNet v1 show only a 1\% difference in the Top 1 and a 0.64\% difference in the Top 5 accuracies.
\end{abstract}

\section{Introduction}
The rapid advancement in artificial intelligence (AI) and machine learning (ML) technologies has spread into various critical domains for our everyday lives, such as healthcare, finance, automotive, and more. 
Among the myriad AI frameworks available, OpenVINO \cite{kozlov2020neural} (Open Visual Inference and Neural Network Optimization) stands out as a powerful open-source toolkit. 
OpenVINO enables developers to optimize and deploy deep learning models on heterogeneous computing platforms, especially targeting edge and embedded devices.

Embedded devices, equipped with neural network accelerators offer great benefits in terms of efficiency and performance. 
However, they also introduce new security challenges, particularly related to hardware attack vectors such as side-channel analysis (SCA) attacks~\cite{batina2022implementation}. 
SCA attacks exploit information leakage from the physical implementation of a system rather than vulnerabilities in the algorithm itself~\cite{mangard2008power}. 
These attacks can reveal sensitive data, such as neural network parameters and architecture, through indirect means such as power consumption, electromagnetic emissions, or execution time analysis~\cite{batina2019csi}.
This complements model stealing/extraction attacks that have been a well-research topic within the past decade~\cite{oliynyk2023know}.
The assumption for these attacks is that the creator of the model invested significant resources into preparing the training data (obtaining the initial data, labeling, and pre-processing it), and also into training the model.
Thus, to have a competitive advantage, they prefer to keep the details of the model confidential.
Apart from that, protecting the model parameters effectively prevents white-box adversarial attacks.

There are two main directions the attacker can take when extracting the model~\cite{oliynyk2023know}: 
\begin{enumerate}
    \item obtaining the exact model parameters,
    \item approximating the model behavior.
\end{enumerate}
When it comes to the first direction, recovering the exact model parameters has been shown possible by fault injection attacks, but only for one layer of a network obtained by transfer learning~\cite{breier2021sniff}.
The goal of model extraction by the means of SCA attacks ultimately leads to the same~\cite{batina2019csi}, however, as the bit-width of the variables used for storing model parameters is generally up to $64$ bits, the noise makes it very hard to determine the exact values from the SCA leakage.
The majority of the literature in this direction thus focuses on approximating the model behavior, generally aimed at creating a substitute model exhibiting similar behavior to the original one.

\vspace{0.5cm}
\noindent
\textbf{Our contribution.}
In this paper, we focus on investigating SCA attacks on neural networks implemented on embedded devices using OpenVINO.
These models are implemented in a quantized manner, thus the model parameters are normally stored in $8-$bit variables, making them more efficient in terms of storage and computational complexity, which is crucial for embedded implementations.
For our experiments, we utilized an OpenVINO implementation of GoogleNet v1 with $\approx 6.7977$ M parameters.
The SCA attack allowed us to reconstruct the model that exhibits very similar accuracies to the original one.
More specifically, the difference in Top $1$ accuracy is $1\%$ and in Top $5$ accuracy it is just $0.64\%$.

\vspace{0.5cm}
\noindent
\textbf{Organization.} 
The rest of the paper is organized as follows.
Section~\ref{sec:background} provides the background on the basic concepts utilized in this paper such as neural networks and SCA attacks, and gives an overview of the related work.
Section~\ref{sec:method} presents the methodology for extracting the model parameters using side channels.
Section~\ref{sec:eval} details our evaluation and highlights the results.
And finally, Section~\ref{sec:concl} concludes this work and provides potential future directions.

\section{Background and Related Work}
\label{sec:background}
\subsection{Neural Networks}
Neural networks (NNs) are a fundamental concept in the field of artificial intelligence and machine learning, inspired by the structure and function of the human brain~\cite{lecun2015deep}.
NNs can recognize patterns, learn from data, and make decisions with minimal human intervention.
They have become important in a wide range of applications, including image and speech recognition, natural language processing, and autonomous systems.
They are supervised ML algorithms -- labeled dataset is used to train their model parameters with an optimization algorithm called (stochastic) gradient descent~\cite{goodfellow2016deep}.

Their structure includes an input layer that receives the initial data for processing, one or more hidden layers (if there are more, we talk about deep neural networks (DNNs)), where the data is transformed through various weights and biases, and an output layer, which produces the result such as classification or prediction.

The original design of an NN, a multilayer perceptron, has been adjusted to fit specific problems.
On the other hand, convolutional neural networks (CNNs) can adaptively learn spatial hierarchies of features, making them effective for image recognition tasks~\cite{li2021survey}.
Another type, recurrent neural networks (RNNs) are capable of capturing temporal dependencies by using cycles within the network, allowing them to analyze sequential data such as time series or natural language.

\subsection{Quantized Neural Networks}
Conventional NNs rely on high-precision floating-point arithmetic, utilizing either $32-$bit or $64-$bit variables.
While such precision helps to achieve state-of-the-art performance, it requires substantial resources and memory bandwidth.
Such a requirement limits their deployment in resource-constrained devices like mobile phones, embedded systems, and edge devices.

Quantized neural networks (QNNs) address this limitation by using lower precision for the weights and activations in the network~\cite{hubara2018quantized}.
QNNs utilize low bit-width numbers such as $8-$bit or $4-$bit integers, or even binary values.
The quantization significantly reduces memory requirements for storing the model and also the amount of computation required, resulting in faster computation and smaller power consumption.

The quantization can either be \textit{uniform}, where all values are evenly distributed across the available range, or \textit{non-uniform} which allows for a more tailored approach based on the data distribution.
Despite the reduced number precision, QNNs often achieve accuracies comparable to their high-precision counterparts, especially when quantization-aware training and tuning are used.

\subsection{OpenVINO Framework}
OpenVINO \cite{kozlov2020neural} is an open-source software toolkit for optimizing and deploying DNN models, mostly on Intel devices but currently also on ARM processors.
It supports a wide range of different platforms from edge to cloud, working with models in TensorFlow, PyTorch, ONNX, TensorFlow Lite, and PaddlePaddle model formats.

In this paper, we also utilize the Neural Network Compression Framework (NNCF)~\cite{kozlov2020neural}.
NNCF provides a suite of post-training and training-time algorithms for optimizing the inference of neural networks in OpenVINO.

\subsection{Side-Channel Analysis Attacks}
SCA is a method originally proposed for recovering secret information from cryptographic implementations~\cite{kocher1999differential}.
Classical cryptanalysis methods focus on analyzing the weaknesses in the algorithms, utilizing techniques such as differential and linear cryptanalysis.
As these are well-studied and understood, the new algorithms are always evaluated against them, rendering cryptanalytic attacks ineffective for a full-round cipher.
However, as of now, there is no cryptographic algorithm inherently resistant to SCAs that analyze the information leakage during the physical operation of the cryptographic device.
Such leakage can come in various forms such as power consumption, electromagnetic emanation, timing of the execution, and even acoustic leakage.

SCAs can recover the secret key very efficiently, sometimes requiring only a single measurement~\cite{banciu2015reliable}.
There are generally two types of attacks: \textit{simple} and \textit{differential}.
The simple attacks only utilize one or a few leakage traces for the attack.
They can be either used as a starting point for the differential attacks, to determine the cipher rounds and round operations in block ciphers, or for a full attack, which is normally the case for public key algorithms~\cite{mangard2008power}.
Differential attacks, on the other hand, utilize higher numbers of measurements and utilize statistical methods to correlate the processed data and the leakage.

Leakage models are used to describe and quantify the information leaked from the implementation under test through side channels during its execution. 
These models help in understanding, analyzing, simulating, and mitigating side-channel attacks. 
The main models used in SCA are as follows:
\begin{itemize}
    \item \textit{Hamming weight model: }assumes that the information leaked is proportional to the number of bits set to \texttt{1} in the data being processed. It is commonly used when the underlying implementation is of software character.
    \item \textit{Hamming distance model: }assumes that the information leaked is proportional to the number of bit transitions (changes \texttt{1}$\to$\texttt{0} or \texttt{0}$\to$\texttt{1}) between two consecutive states of the data. 
    \item \textit{Stochastic model: }goes further than the previous two and assumes that every bit of the analyzed variable leaks differently. 
\end{itemize}
We give a more thorough explanation of leakage models in Section~\ref{sec:leakage_models}.

\subsection{Related Work -- SCAs on Neural Networks}
In the first part of this section, we will focus on various attacks on NNs, in the second part we will outline some of the possible defenses.

\subsubsection{Attacks.}
The first comprehensive work in this area by Batina et al.~\cite{batina2019csi} explored how neural networks can be reverse-engineered using electromagnetic (EM) analysis. 
The paper demonstrated that a passive, non-invasive attacker can extract model details such as activation functions, number of layers, neurons, output classes, and weights from neural networks by analyzing EM signals. 
The experiments were conducted on an ARM Cortex-M3 microcontroller, showing the feasibility of such attacks on widely used hardware platforms.

Wang et al.~\cite{wang2023side} investigated the vulnerability of in-memory computing (IMC) systems to side-channel attacks. 
By simulating power traces of IMC macros, the researchers demonstrated that attackers can reverse-engineer neural network models, extracting details like layer types and convolution kernel sizes without prior knowledge. 

Gao et al.~\cite{gao2023deeptheft} presented a novel attack method called DeepTheft, which targets DNN models deployed in Machine Learning as a Service (MLaaS) environments. 
By exploiting the Running Average Power Limit (RAPL)-based power side channel, the work demonstrated that it is possible to accurately recover complex DNN model architectures, including detailed layer-wise hyperparameters, even with low sampling rates. 

Ryu et al.~\cite{ryu2023gamma} introduced a novel attack method called Gamma-Knife. 
This attack leverages software-based power side channels to extract the architecture of neural networks without requiring physical access or high-precision measuring equipment. 
By utilizing statistical metrics, the Gamma-Knife attack can accurately determine key architectural details such as filter size, depth of convolutional layers, and activation functions. 
The researchers demonstrated the effectiveness of this attack on popular neural networks like VGGNet, ResNet, GoogleNet, and MobileNet, achieving an accuracy of approximately $90\%$.

Nagarajan et al.~\cite{nagarajan2023scann} investigated the vulnerabilities of spiking neural networks (SNNs) to power SCAs. 
The authors demonstrated that different synaptic weights and neuron parameters in SNNs produce distinct power and spike timing signatures, making them susceptible to SCAs. 
Through eight unique attacks, they showed that an adversary can reverse-engineer the specifications of an SNN.

\subsubsection{Countermeasures.}

Dubey et al.~\cite{dubey2020maskednet} proposed a novel hardware design that incorporates masking techniques. 
This design includes masked adder trees for fully connected layers and masked Rectifier Linear Units for activation functions. 
Experiments on a SAKURA-X FPGA board show that the proposed protection significantly increases the latency and area cost but effectively mitigates first-order differential power analysis attacks.

In \cite{dubey2022modulonet}, the authors proposed using modular arithmetic to make neural networks more compatible with masking techniques. 
They demonstrated this approach on binarized neural networks (BNNs) and developed novel masking gadgets using Domain-Oriented Masking (DOM). 
Their implementation on an FPGA showed that this method can achieve similar latency while reducing flip-flop (FF) and lookup table (LUT) costs by $34.2\%$ and $42.6\%$, respectively, compared to state-of-the-art protected implementations. 
They verified the first-order side-channel security of their design with up to 1 million traces.

Breier et al.~\cite{breier2023desynchronization} proposed a method to protect neural networks from side-channel attacks by making the timing analysis of activation functions more difficult. 
The authors introduced a desynchronization technique that adds random delays to the computation of activation functions, effectively hiding the dependency on the input and activation type. 
They experimentally verified the effectiveness of this countermeasure on a 32-bit ARM Cortex-M4 microcontroller, showing that it significantly reduces side-channel information leakage. 
The overhead of this method varies depending on the number of neurons, with an example overhead of $2.8\%$ to $11\%$ for a fully connected layer with $4,096$ neurons in the VGG-19 network

As the topic of SCA on NNs is comprehensive and the number of papers in this area increases every year, we suggest interested readers to explore one of the recently published surveys for a full overview of the state-of-the-art~\cite{batina2022implementation,chabanne2021side,mendez2021physical}.

\section{Methodology}
\label{sec:method}
\subsection{Binary Representation of Numbers in Memory}
When a value $\boldsymbol{v}$ is processed by a computational device, it is represented as a binary string.
For integer values, $\boldsymbol{v}$ is encoded using two's complement representation.
For floating-point values, the IEEE 754 standard is employed.

For example, the two's complement representations for integers between $-8$ and $7$ are shown in Table \ref{tab:twoscomp}.

\begin{table}[b]
    \centering\small
    \begin{tabular}{|c|c||c|c|}\hline
     7 & 0111 & -1 & 1111\\
     6 & 0110 & -2 & 1110\\
     5 & 0101 & -3 & 1101\\
     4 & 0100 & -4 & 1100\\
     3 & 0011 & -5 & 1011\\
     2 & 0010 & -6 & 1010\\
     1 & 0001 & -7 & 1001\\
     0 & 0000 & -8 & 1000\\
     \hline
    \end{tabular}
    \vspace{3mm}
    \caption{Two's complement encoding for integers between $-8$ and $7$.}
    \label{tab:twoscomp}
\end{table}

A $32$-bit (single precision) floating-point representation, following the IEEE 754 standard, comprises $1$ sign bit, $8$ exponent bits, and $23$ fraction bits (also known as the significand or mantissa).
The sign bit indicates the number's sign: $0$ for positive and $1$ for negative.
The exponent is encoded using a biased representation with a bias of $127$, enabling the exponent to represent both positive and negative values.
The actual exponent is calculated by subtracting $127$ from the $8$-bit exponent.
The mantissa encodes the significant digits of the number, incorporating an implicit leading bit of $1$, which is assumed but not stored.
Specifically, the binary string $b_{31}b_{30}\cdots b_0$ represents the integer given by
\[
(-1)^{b_{31}}\times 2^{b_{30}b_{29}\dots b_{23}-127}\times1.b_{22}b_{21}\dots b_0,
\]
where the exponent $b_{30}b_{29}\dots b_{23}$ represents the integer given by
\[
b_{30}2^{8}+b_{29}2^{7}+\dots +b_{23}
\]
and $1.b_{22}b_{21}\dots b_0$ represents the number
\[
1+\frac{b_{22}}{2}+\frac{b_{21}}{2^{2}}\dots+\frac{b_0}{2^{23}}.
\]
For example, the binary string 
\[
01000001001101100000000000000000
\]
has sign bit $0$, exponent
\[
10000010=130
\]
and mantissa 
\[
01101100000000000000000.
\]
This mantissa represents
\[
1.011011_2=1+\frac{1}{4}+\frac{1}{8}+\frac{1}{32}+\frac{1}{64}=1.421875.
\]
The number the binary string represents is then given by
\[
(-1)^0\times2^{130-127}\times1.421875=11.375.
\]

\subsection{Leakage Models}
\label{sec:leakage_models}
In SCA, the leakage model characterizes the relationship between the computational leakage and the secret value $\boldsymbol{v}$ processed by the device.
Suppose 
\[
\boldsymbol{v}=v_{n-1}v_{n-2}\dots v_1v_0\in\FF_2^n
\]
is represented as a binary string of length $n$ in the computer memory.
The \textit{Hamming weight} of $\boldsymbol{v}$, denoted HW$(\boldsymbol{v})$, is given by the number of bits that are equal to $1$ in $\boldsymbol{v}$.
For example,
\[
\text{HW}(110)=2,\quad \text{HW}(000)=0,\quad \text{HW}(110101)=4.
\]

Let $\varepsilon \sim \mathcal{N}(0, \sigma^2)$ denote the noise in the leakage, modeled as a normal distribution with mean $0$ and variance $\sigma^2$.

A \textit{Hamming weight leakage model} \cite[Section 4.2]{hou2024cryptography} suggests that the leakage $\mathcal{L}(\boldsymbol{v})$ is given by
\[
\mathcal{L}(\boldsymbol{v})=\text{HW}(\boldsymbol{v})+\varepsilon.
\]
A \textit{stochastic leakage model} \cite[Section 4.3]{hou2024cryptography} expresses the leakage as
\begin{equation}\label{eq:stochastic}
    \mathcal{L}(\boldsymbol{v})=\sum_{s=0}^{n-1}\alpha_sv_s+\varepsilon,
\end{equation}
where $\alpha_s$ ($s=0,1,\dots, n-1$) are real numbers.
These numbers $\alpha_s$ are referred to as the \textit{coefficients} of the stochastic leakage model.

\subsection{Correlation Power Analysis}
We employ the correlation power analysis (CPA) methodology originally devised for cryptographic implementation attacks \cite{mangard2008power}.
CPA focuses on calculating Pearson's correlation coefficient between observed leakages and hypothetical leakages derived from a guessed secret value.
The correct secret value is anticipated to yield the highest correlation coefficient, indicating a closer match between actual and modeled leakages.

Next, we outline the application of this approach to recover the secret weight and bias values of QNNs.
Let $w$ denote a secret value, and let
\[
w_1,w_2,\dots,w_N
\]
be all the possible values of the secret weight.
For $8-$bit QNNs, 
\[
w_1,w_2,\dots,w_N
\]
are given by
\[
-128,-127,-126,\dots,-2,-1,0,1,2\dots,126,127
\]
and $N=256$.
During the inference phase, the weight value $w$ is multiplied with a neuron input, and the resulting product is utilized for subsequent computations.
To recover the value of $w$, our focus lies on this multiplication operation.
We assume the attacker has knowledge of the neuron input:
\begin{itemize}
    \item When $w$ belongs to the first hidden layer, the neuron input can be inferred from the NN input, which is known to the attacker.
    \item For inner layers, we anticipate the attacker first recovering parameters from preceding layers and subsequently computing the input of the inner layer using the recovered parameters.
\end{itemize}
Let $x$ represent the neuron input multiplied by $w$, yielding the product $\boldsymbol{v}$.

In our experiments, we simulate actual leakage using stochastic leakage models with different coefficients. We assume that the attacker either uses the Hamming leakage model or possesses profiling capabilities to correctly identify these coefficients for the stochastic leakage model.

For the attack procedure, we first provide the network with randomly generated $M$ inputs to obtain random neuron inputs $x$, denoted by
\[
x_1,x_2,\dots,x_M.
\]
Subsequently, the leakage associated with the product $\boldsymbol{v}$ is simulated according to the predefined leakage model.

Let $l_j$ denote the leakage corresponding to the input value $x_j$ for $j = 1, 2, \dots, M$.
In this case, $l_j$ simulates the leakage of
\[
\boldsymbol{v}_j:=x_j\times w.
\]
The simulation follows a stochastic leakage model (Equation \ref{eq:stochastic}) with pre-defined coefficients and random noise.

Next, we make \textit{hypotheses} of the value $\boldsymbol{v}$:
For each possible value, or \textit{hypothesis}, of $w$, $w_1,w_2,\dots,w_N$ and each input value of $x$, $x_1,x_2,\dots, x_M$, we compute the hypothetical value of the product:
\[
\hat{\boldsymbol{v}}_{ij}=w_i\times x_j.
\]
We then calculate the hypothetical leakage, $\mathcal{H}_{ij}$, of this hypothetical product $\hat{\boldsymbol{v}}_{ij}$ using the attacker's leakage model. This leakage model can either be the Hamming weight model or a profiled stochastic leakage model, assumed to match the one used for simulating leakages.

For example, if the Hamming weight leakage model is employed for the attack, the hypothetical leakage of $\hat{\boldsymbol{v}}_{ij}$ is given by
\[
\mathcal{H}_{ij}=\text{HW}(\hat{\boldsymbol{v}}_{ij}).
\]

With CPA, we compute the Pearson correlation between the simulated leakage and the hypothetical leakage for each weight hypothesis using the formula:
\[
r_i := \frac{\sum_{j=1}^{M}(\mathcal{H}_{ij}-\overline{\mathcal{H}_i})(l_j-\overline{l})}{\sqrt{\sum_{j=1}^{M}(\mathcal{H}_{ij}-\overline{\mathcal{H}_i})^2}\sqrt{\sum_{j=1}^{M}(l_j-\overline{l})^2}},\quad i=1,2,\dots,N.
\]
Here, 
\[
\overline{\mathcal{H}_i}=\frac{1}{M}\sum_{j=1}^M \mathcal{H}_{ij}
\]
is the averaged hypothetical leakage for weight hypothesis $w_i$, and
\[
\overline{l}=\frac{1}{M}\sum_{j=1}^M l_j
\]
is the averaged simulated leakage.

In case the hypothesis of the weight value, $w_i$, is correct, we expect the corresponding coefficient $r_i$ to have a high absolute value.
Our guessed weight is then given by
\[
w_{\text{guessed}}=w_{i_0},\quad \text{where}\quad i_0=\argmax_{i} |r_i|.
\]
Furthermore, in case multiple weight hypotheses achieve the highest absolute value of $r_i$, we select the one with the smallest absolute difference between the hypothetical and simulated leakages:
\[
w_{\text{guessed}}=w_{i_0},
\]
where
\[
i_0=\argmin_{i: w_i \text{ achieves the highest absolute value of }r_i}\left\lvert \sum_{j=1}^M (\mathcal{H}_{ij}-l_j )\right\rvert.
\]

An algorithmic representation of the described procedure using Hamming weight leakage model for the attack is provided in Algorithm \ref{alg:weight}.

The attack to recover bias values is similar.
Instead of focusing on the multiplication operation, we concentrate on the addition operation, which adds the bias to the product of weights and neuron inputs.
We assume the attacker has already recovered the weight values in the given layer.
With knowledge of the neuron inputs, the attacker can compute the value added to the secret bias.

\begin{algorithm}
\KwIn{$\alpha_s$ $(s=0,1,\dots,32)$,  $w$, $w_1,w_2,\dots,w_N$, $M$, $\mu$, $\sigma^2$
\tcp{$\alpha_s$ represents the coefficients of the stochastic leakage model, $w$ denotes the correct weight value, and $M$ is the number of random inputs to be generated. 
$w_1,w_2,\dots,w_N$ are all the possible values of the weight.
The parameters $\mu$ and $\sigma^2$ correspond to the mean and variance, respectively, of the normal distribution used to model noise.}
}
\KwOut{recovered weight}
randomly generate $M$ inputs $x_1, x_2,\dots, x_M$\;
\textbf{array} of size $M\quad$ $l$\tcp{array to store simulated leakages}
\For{$j=1,2,\dots,M$}{
$\boldsymbol{v}=x_j\times w$\tcp{compute the correct product}
generate noise $\varepsilon$\tcp{generate noise from a normal distribution with mean $\mu$ and variance $\sigma^2$}
$l[j-1]=\mathcal{L}(\boldsymbol{v})$\tcp{simulate leakage using the coefficients $\alpha_s$ and noise $\varepsilon$ following Equation \ref{eq:stochastic}}
}
$\overline{l}=\displaystyle\frac{1}{M}\sum_{j=0}^{M-1} l[j]$\tcp{the averaged leakage}
\textbf{array} of size $N\quad$ $r_{abs}$\tcp{array to store absolute values of correlation coefficients}
\textbf{array} of size $N\quad$ $\texttt{dif}_{abs}$\tcp{array to store absolute values of differences between hypothetical leakages and the simulated leakages}
\For{$i=1,2,\dots,N$}{
\textbf{array} of size $M\quad$ $\hat{\boldsymbol{v}}$\tcp{array to store hypothetical product values}
\textbf{array} of size $M\quad$ $\mathcal{H}$\tcp{array to store hypothetical leakages}
\For{$j=1,2,\dots,M$}{
$\hat{\boldsymbol{v}}[j-1]=w_i\times x_j$\tcp{the hypothetical product for hypothesis of the weight value $w_i$ and input $x_j$}
$\mathcal{H}[j-1]=\text{HW}(\hat{\boldsymbol{v}}[j-1])$\label{line:alg:leakagemodel}\tcp{the hypothetical leakage for hypothesis of the weight value $w_i$ and input $x_j$}
}
$\overline{\mathcal{H}}=\displaystyle\frac{1}{M}\sum_{j=0}^{M-1} \mathcal{H}[j]$\tcp{averaged hypothetical leakage}
$r_{abs}[i-1] = \displaystyle\frac{\sum_{j=0}^{M-1}(\mathcal{H}[j]-\overline{\mathcal{H}})(l[j]-\overline{l})}{\sqrt{\sum_{j=0}^{M-1}(\mathcal{H}[j]-\overline{\mathcal{H}})^2}\sqrt{\sum_{j=0}^{M-1}(l[j]-\overline{l})^2}}$\tcp{correlation coefficient for weight hypothesis $w_i$}
$r_{abs}[i-1]=\lvert r_{abs}[i-1]\rvert$\tcp{compute the absolute value of the correlation coefficient}
$\texttt{dif}_{abs}[i-1]=\left\lvert\displaystyle\sum_{j=0^{M-1}}(\mathcal{H}[j]-l[j])\right\rvert$\tcp{the absolute value of the sum of the differences between the hypothetical and simulated leakage values}
}
\caption{A simulated attack for recovering a single weight using the Hamming weight model}
\label{alg:weight}
\end{algorithm}

\begin{algorithm}
  \LinesNumbered
\setcounter{AlgoLine}{19}
$\texttt{max\_cor}= \max(r_{abs})$\tcp{find the maximum value in $r_{abs}$}
$\texttt{ind}=\text{indices of \texttt{max\_cor} in } r_{abs}$\tcp{find all weights that achieve the highest absolute correlation coefficient}
\If{len(\texttt{ind})$==1$}{
\Return $w_{\texttt{ind}[0]}$\tcp{if there is just one weight that achieve the highest absolute value of correlation coefficient, we take it to be the guessed weight}
}
\Else{
$\texttt{min\_dif}=\min \texttt{dif}_{abs}$\tcp{find the minimum value in $\texttt{dif}_{abs}$}
$\texttt{ind2}=\text{index of \texttt{min\_dif} in }\texttt{dif}_{abs}$\tcp{in case multiple weight hypotheses achieve the highest absolute value of correlation coefficient, we select the one with the smallest absolute difference between the hypothetical and simulated leakages}
\Return $w_{\texttt{ind2}[0]}$
}
\end{algorithm}

\section{Evaluation}

In this section, we will report the results for our experimental evaluation. We conducted a simulation for weight and bias recovery and presented the results. We then extended the work and report the findings for full QNN model.

\label{sec:eval}
\subsection{Evaluation Scenarios}
We first conducted the experiment on the recovery of weight and bias in the neural network.
For the experimental evaluation, the leakage simulations were conducted employing stochastic leakage models with varying bit coefficients $\alpha_s$ (see Equation \ref{eq:stochastic}).

We considered two distinct leakage settings, each with different bit coefficients for the stochastic leakage model. 
Additionally, we examined two attack settings: the first assumes that the attacker does not have the full capability to profile the device and thus uses a Hamming weight leakage model, which may be inaccurate in this context; the second assumes that the attacker could profile the exact bit coefficients for the stochastic leakage model.

Preliminary tests were conducted on several boards to profile the leakage behavior. 
Upon profiling, we observed the common variances of the bit coefficients for the stochastic leakage model, which will then be used and adopted in our experiments. 
The following three scenarios were considered:

\begin{itemize}
    \item \textbf{Scenario 1}: The bit coefficients for the stochastic leakage model are randomly generated from a normal distribution with mean $1$ and variance $0.09$. 
    In this setting, there are deviations between each bit coefficient. However, they are closer to the Hamming weight leakage model.
    Here, we consider an attacker who cannot profile the device and use a Hamming weight leakage model for recovery of the weights and biases.
    \item \textbf{Scenario 2}: The bit coefficients for the stochastic leakage model are randomly generated from a normal distribution with mean $1$ and with a higher variance value of $1$. Similarly, for the attack, the attacker does not know the exact bit coefficient values and assumes a Hamming weight leakage model.
    \item \textbf{Scenario 3}: The bit coefficients for the stochastic leakage model are randomly generated from a normal distribution with mean $1$, with a similar higher variance of $1$. 
    In this scenario, the attacker is assumed to have accurately profiled the leakage and identified the correct bit coefficient values and, as such, can build a more precise leakage model for the key recovery attack.
\end{itemize}

After setting the leakage model bit coefficients, we generate the simulated side-channel traces. 
For our experiment, our main focus will be on investigating the impact of varying bit coefficients of the leakage model.
As such, we will be fixing the noise level. The noise in the leakage model, $\varepsilon \sim \mathcal{N}(0, 0.5)$, follows a normal distribution with mean $0$ and variance $0.5$. 
This ensures a fair comparison of all the attack scenarios.

Using this setting, we generate a set of simulated traces. We simulate weight and bias recovery using CPA with $100,000$ traces for one simulated attack, i.e. $M=100,000$ following the notations from Section \ref{sec:method}.
Specifically, for a weight recovery attack, we set $\mu=0$ and $\sigma^2=0.5$ in Algorithm \ref{alg:weight}.
In Scenario 1, $\alpha_s$ values were drawn from a normal distribution with a mean of $0$ and a variance of $0.09$.
For Scenarios 2 and 3, $\alpha_s$s values were generated from a normal distribution with a mean of $0$ and a variance of $1$.
Weight recovery in Scenarios 1 and 2 follows Algorithm \ref{alg:weight} while Scenario 3 adheres to the algorithm with one modification: line \ref{line:alg:leakagemodel} is replaced by
\[
\mathcal{H}[j-1]=\mathcal{L}(\hat{\boldsymbol{v}}[j-1]),
\]
in accordance with Equation \ref{eq:stochastic} using the pre-generated leakage coefficients and noise.

For the recovery of weights, a total of $250$ attacks were conducted, with randomly generated weight values for each attack. 
Similarly, $250$ attacks were simulated for the recovery of biases, with randomly generated bias values for each attack.

We record the \textit{accuracy} as the rate at which the correct weight or bias is successfully recovered. We denote the average error as the absolute difference between the actual weight or bias and the recovered value.
The results for each weight and each bias under different scenarios are reported as follows:
\begin{itemize}
    \item Scenario 1: We achieved accuracy $80.7\%$ with average error $1.86$ for weight recovery and accuracy $29.6\%$ with error $0.19$ for bias recovery.
    \item Scenario 2: We achieved accuracy $33.5\%$ with average error $1.04$ for weight recovery and accuracy $7.6\%$ with error $26014.7$ for bias recovery.
    \item Scenario 3: We achieved accuracy $46.7\%$ with average error 2 for weight recovery and accuracy $100\%$ for bias recovery.
\end{itemize}

As observed from the results, when the variance of the bit coefficient is low, the Hamming weight leakage model proves sufficient, as indicated by the high accuracy in weight recovery. For bias recovery, although the accuracy is not high, the low average error can be attributed to targeting the addition operation rather than the multiplication operation, as is the case with weight recovery. 

Interestingly, when the attacker can profile the bit coefficients, even with high variance, the accuracy for weight recovery decreases, but perfect accuracy is achieved for bias recovery. Nonetheless, for high variance in the bit coefficient, the use of the Hamming weight leakage model results in poor performance.

\subsection{Application to Full Networks}
\label{sec:application}
After analyzing the results for weight and bias recovery, we would like to evaluate the attack on full network.
To evaluate the attack's performance on complete networks, we simulated the recovery of existing quantized neural networks and assessed the test accuracy. 

\begin{figure*}[tb]
    \centering
    \includegraphics[width=0.8\linewidth]{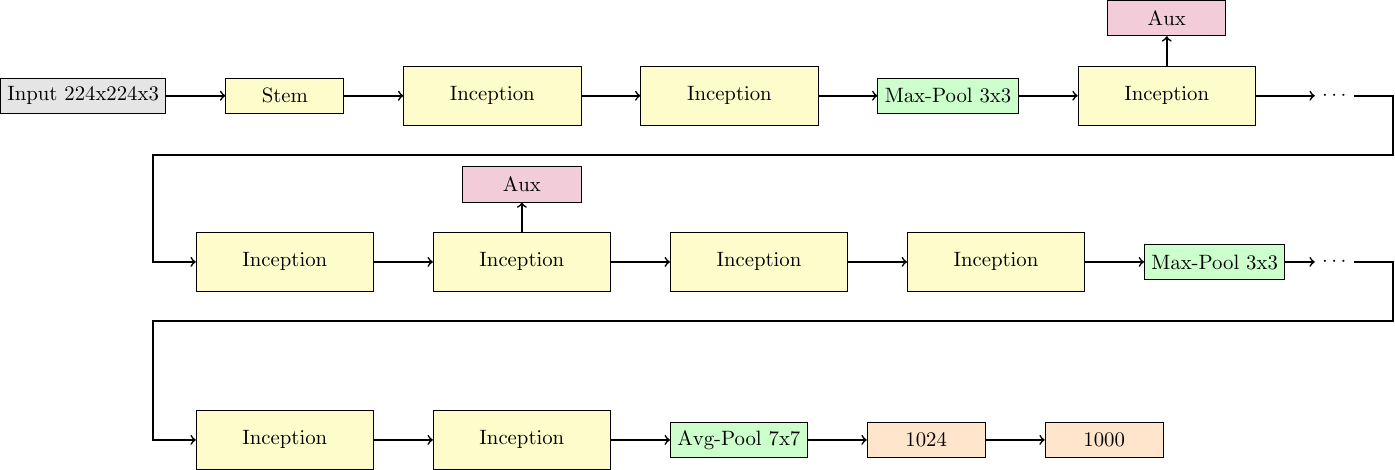}
    \caption{Architecture of GoogleNet v1.}
    \label{fig:googlenet}
\end{figure*}

As the target model, we consider quantized GoogleNet v1 model~\cite{szegedy2015going} from OpenVINO Model Zoo\footnote{\url{https://github.com/openvinotoolkit/nncf/blob/develop/docs/ModelZoo.md}}, on the ImageNet dataset.
GoogleNet, also known as Inception v1, is a deep convolutional neural network that was developed by Google for the ImageNet Large-Scale Visual Recognition Challenge (ILSVRC) 2014. 
It introduced the Inception module, which allows the network to capture multi-scale features while maintaining computational efficiency.
The network has $\approx6.7977$ M parameters, its high-level architecture is depicted in Figure~\ref{fig:googlenet}. As such, it is a decent target for evaluating our model extraction approach.

ImageNet is a large-scale visual database designed for use in visual object recognition research~\cite{deng2009imagenet}. 
It contains over $14$ million images that have been hand-annotated to indicate the objects present in them. 
These images are organized according to the WordNet hierarchy, which includes more than $20,000$ categories, such as ``balloon'' or ``strawberry.''

We considered the recovery of the quantized GoogleNet v1 model under 3 scenarios presented earlier. For each scenario, we repeat the procedure 5 times. We reported the results in Table~\ref{tab:acc_sa_model}. We report the top 1 and top 5 accuracy (\%) using the accuracy checker from OpenVINO. As a comparison, we also reported the same accuracies for the original GoogleNet v1 model.

\begin{table}
\centering
\begin{tabular}{lcc} %
\hline
   Model & Top 1 Acc. (\%) & Top 5 Acc. (\%) \\
\hline
\hline
  GoogleNet v1 (Original) & 62.36 & 84.91\\
  GoogleNet v1 Scenario 1 & 61.36 & 84.27\\
  GoogleNet v1 Scenario 2 & 0.10 & 0.50\\
  GoogleNet v1 Scenario 3 & 59.29 & 84.74\\
\hline
\end{tabular}
\caption{Top 1 and Top 5 Accuracy for Original and Reconstructed Model on ImageNet dataset}
\label{tab:acc_sa_model}
\end{table}%

As can be seen from Table~\ref{tab:acc_sa_model}, we observe that the recovered model from Scenario 1 and 3 performs almost similarly to the original GoogleNet model. This makes sense since the leakage model quite matches the leakage behavior. However, for Scenario 2, when the attacker is using the Hamming weight leakage model when targeting leakage with high variance for their bit coefficients, the performance degrades significantly. 
This follows with the idea of the scenario which shows the worst case for the attacker, with random coefficients and no knowledge of them\footnote{However, such a scenario would be rare in the real world.}.

In this case, the observed results indicate that the leakage model plays a crucial role in the success of the model recovery. As an attacker, the main target is then to develop a leakage model closer to the actual leakage. However, this task is not trivial, and without profiling, the attacker can only rely on the standard Hamming weight model. As such, solving this issue will be a main priority for future work.

\section{Conclusion}
\label{sec:concl}
In this paper, we presented an SCA attack on quantized neural network models implemented in the OpenVINO framework.
These models are meant to be deployed in embedded devices, thus allowing the attacker physical access to realize the side-channel measurements.
Moreover, the quantization, restricting the model variables to be stored in $8-$bit variables, greatly increases the precision of the parameter recovery.
Our results show that for GoogleNet v1, the top model recovered by the SCA attack achieves similar accuracy to the original model, differing by $1\%$ in Top $1$ and by only $0.64\%$ in Top $5$ accuracies.

\subsection*{Acknowledgement}
OpenAI’s ChatGPT-4 was used to improve the clarity and readability of this manuscript. After using this tool, the authors reviewed and edited the content as needed and take full responsibility for the manuscript's content.
This project has received funding from the European Union's Horizon 2020 Research and Innovation Programme under the Programme SASPRO 2 COFUND Marie Sklodowska-Curie grant agreement No. 945478.
This research is funded by the European Commission, under the Horizon Europe project aerOS, grant number 101069732.

This research is supported by the National Research Foundation, Singapore, and Cyber Security Agency of Singapore under its National Cybersecurity Research \& Development Programme (Development of Secured Components \& Systems in Emerging Technologies through Hardware \& Software Evaluation < NRF-NCR25-DeSNTU-0001 >). Any opinions, findings and conclusions or recommendations expressed in this material are those of the author(s) and do not reflect the view of National Research Foundation, Singapore and Cyber Security Agency of Singapore.

\bibliographystyle{plain}
\bibliography{bib}

\end{document}